\documentstyle[12pt]{article}

\def\air{\rightarrow}
\def\p{$p$\--adic~}
\def\Q{{\bf Q}}
\def\Z{{\bf Z}}
\def\M{{\cal M}}
\def\B{{\cal B}}
\def\E{{\cal E}}
\def\R{{\bf R}}

\begin{document}
\begin{flushright}
EFI 92-11 \\
\end{flushright}

\begin{center}
$Z_n$ Baxter Models and Quantum Symmetric Spaces\footnote{Work
supported in part by the NSF: PHY-9000386} \\

\end{center}
\bigskip
\medskip

\begin{center}
Peter G. O. Freund and Anton V. Zabrodin\footnote{Permanent address:
Institute of Chemical Physics, Kosygina Str.~4, SU-117334, Moscow, Russia.} \\

\smallskip
{ \em Enrico Fermi Institute, Department of Physics\\
and Mathematical Disciplines Center \\
University of Chicago, Chicago, IL 60637 \\ }

\end{center}

\bigskip
\medskip
\centerline{ABSTRACT}
\begin{quote}

The scattering of two excitations (both of the simplest kind) in the
magnetic model related to the $Z_n$\--Baxter model is naturally described for
$n \air \infty$ in terms of the Macdonald polynomials for root system $A_1$.
These polynomials play the role of zonal spherical functions for a two
parameter family of quantum symmetric spaces.
These spaces ``interpolate'' between various $p$\--adic and real symmetric
spaces.

\end{quote}

\vfill
\eject

Work [1]-[5] on the theory of spherical functions, on scattering phenomena
and on strings has recently converged on the idea that there may exist a
two\--parameter family of ``quantum'' symmetric spaces that ``interpolate''
between real and \p symmetric spaces.
In this paper we wish to establish this
``$p$\--adics\--quantum\--group\--connection''.
Our procedure for doing this is rooted in the Physics of certain integrable
models, the $Z_n$\--Baxter models of statistical mechanics and their field
theoretic counterparts.
We shall therefore be able to shed new light on this Physics.
In particular the problem of finding the $S$\--matrix for the scattering of
two (simplest) excitations in such models will be reformulated as that of
finding the zonal spherical functions of a certain (elliptic) quantum
symmetric space.

To fix ideas, let us start with some concrete and simple examples.
Consider a point particle, moving on the 2\--sphere $S^2 = SU(2)/SO(2)$.
Its wave function is an eigenfunction of the laplacian on the compact real
symmetric space $S^2$.
If we restrict ourselves to states invariant under $SO(2)$ rotations, say
around the $z$\--axis, then the eigenfunctions will be
the zonal spherical functions  (zsf's) on $S^2$,
the familiar Legendre
polynomials $P_n (\cos \theta )$ in the continuous ``coordinate'' $\cos
\theta$.
The corresponding energy eigenvalues $E_n = n (n+1)$ define a discrete
spectrum, since $S^2$ is compact,
$SU(2)$ being a compact real form of $SL(2)$.
Were we,  on the other hand,
to start from a  non\--compact and  not real
``cognate'' of the sphere, the  \p hyperbolic plane $H_p = SL(2, \Q_p )/ SL
(2, \Z_p ) ~ (
\Q_p =$ field of \p numbers, $\Z_p = $ ring of \p integers),
then the wave functions
of a point particle
would be [6] the
zsf's of $H_p$, the Mautner\--Cartier [7], [8] polynomials
$\rho_k (n)$.
Here $n$ is the {\em quantized} distance on $H_p$, and $k$ is related to the
eigenvalue of the laplacian on $H_p$, whose spectrum is continuous, $SL(2,
\Q_p)$ being
non\--compact.
As a ``compensation'' the coordinate $n$ is now quantized, the \p hyperbolic
plane being a {\em discrete} space [9], a
Cayley\--Bethe\--Bruhat\--Tits\--tree with $p+1$ edges incident at each
vertex.
In switching from $S^2$ to $H_p$ two things have changed: the nature of the
field from which the entries of the $2 \times 2$ matrix are picked has
changed from $\R$ to $\Q_p$ and the compactness has been lost.
Otherwise the problem remained the same:
find the zonal spherical functions of the (real) 2\--sphere in one case,
and of the \p hyperbolic planes in the other cases.
One may wonder whether a more general structure could lie ``above'' all
these problems, which when specialized $--$ say, by fixing some parameters $--$
can be brought to reduce to each and every one of these symmetric spaces:
the real sphere and the \p hyperbolic planes.
As we already noted, earlier work [1]-[5] has conjectured (based on ample
partial evidence) that the ``agency'' which interpolates between all these
symmetric spaces is a quantum symmetric space with two deformation
parameters.
We then expect that as we vary the two deformation parameters of the
conjectured quantum symmetric space, we shall,
in particular, ``land'' on each and every one
of the ordinary symmetric spaces $S^2 , H_p$.
In string theory any of these quantum symmetric spaces can be thought of as
a candidate string world\--sheet [3].
Mathematically, all this is one special case of a generic situation that
occurs for any ``admissible pair'' of root systems [1], [2].

We shall settle this conjecture in the affirmative.
We shall show that
the quantum symmetric spaces constructed from
the elliptic quantum algebra of Sklyanin [10], [11] and
its generalizations [12]-[14]
have the ``interpolating'' property called for in the conjecture.
The elliptic quantum algebras thus provide a conceptual understanding of the
extensive evidence on the basis of which this ``$p$\--adics\--quantum group
connection'' was conjectured [5]-[9].

In this paper we shall only sketch proofs, emphasizing however conceptual
matters.
The full details are to be found in an expanded version of this
paper [15].

Let us start from the Mautner\--Cartier polynomials [7], [8].
If $E$ is the eigenvalue of the Laplacian on the $H_p$\-- tree, it is
convenient to introduce a momentum\--like variable $k$, by
$$
E = p^{1/2} (p^{ik/2} + p^{-ik/2} ) -p - 1
\eqno(1.a)
$$
in terms of which the Mautner\--Cartier polynomials are
$$
\rho_k (n) =
\frac{p}{p+1} \left \{
c_p (ik) p^{(ik-1){n \over 2}} + c_p (-ik)p^{(-ik-1){n \over 2}} \right \}
\eqno(1.b)
$$
with
$$
c_p(s) =
\frac{\zeta_p (s)}{\zeta_p (s+1)} ~~, ~~~
\zeta_p(s) =
\frac{1}{1-p^{-s}} ~~.
\eqno(1.c)
$$
These $\rho_k (n)$ are polynomials in the variable ${1 \over 2}(p^{ik/2} +
p^{-ik/2})$, hence their name.
Recall further that the discrete variable $n$ is the radial coordinate on
the tree.
What matters for us, is that in Eq.~(1b) we clearly recognize an outgoing
wave and an incoming wave, so that their ($n$\--independent) coefficients
$c_p(+ik)$ and $c_p (-ik)$ are [1] the Jost functions of the ``scattering''
problem on the tree.
In terms of them the $S$\--matrix is given by
$$
S_p (k) =
\frac{c_p (ik)}{c_p(-ik)}
\eqno(2)
$$
This Jost function $c_p(ik)$ $--$ or {\em Harish\--Chandra c\--function}
[16] as it is known in the theory of symmetric spaces $--$ thus plays a
double r\^{o}le.
On the one hand it determines the scattering matrix, on the other hand it
calls for the Mautner\--Cartier polynomials and thus identifies the
symmetric space $H_p$.
It is this double r\^{o}le of the Harish\--Chandra $c$\--function that we
shall exploit to find the two\--parameter family of  quantum symmetric
spaces that interpolate between the real two\--sphere and the \p hyperbolic
planes.

We shall not deal directly with the quantum or classical symmetric spaces
themselves, but only with their zonal spherical functions (zsf's).
A two\--parameter family of orthogonal polynomials which interpolate between
the Mautner\--Cartier polynomials (the zsf's of $H_p$) and the Legendre
polynomials (the zsf's of $S^2$) has been known for a long time.
It is the family of Rogers\--Askey\--Ismail (RAI) polynomials [17], [18], or
equivalently the family of Macdonald polynomials [1], [2] for root system
$A_1$.
 From the asymptotics of these RAI polynomials one can extract [1], [2], [15]
a Harish\--Chandra\--like $c$\--function $c(iu; t|q)$ and it is this
function we shall find in the r\^{o}le of Jost function for the scattering
of two excitations in a physical system with an underlying elliptic Sklyanin
algebra.

The system we have in mind is the $Z_n$\--Baxter model $\B_n$ [19], [20]
and the related
generalized magnetic model $\M_n$ [21] ($\B_2$ is the original Baxter
8\--vertex model, $\M_2$ the XYZ model, we shall also be interested in the
$n \rightarrow \infty$ limit).
The quantum algebra which allows a solution of the
Yang\--Baxter equations of these models, thus leading to the existence of
infinitely many commuting operators and therefore to the
integrability of the $Z_n$\--Baxter models, has been studied by Sklyanin
[10], [11], Cherednik [12], [13] and by Odeskii and Feigin [14].
For the $\B_n$ or $\M_n$ model this algebra is $Q_{n^2, n-1} ~ ( \E ,
\gamma)$ in the notation of [14], we shall simply call it $Q_n$.
It's data are the integer $n$, an elliptic curve $\E$ and a point $\gamma$
on $\E$.
In particular $Q_2 \equiv Q_{4,1} ( \E, \gamma )$ is the original elliptic
Sklyanin algebra of $\B_2$ or $\M_2$.
The appearance of an elliptic curve is hardly surprising in view of the
Baxter\--Belavin parametrization [19], [20] of the $n^2 \times n^2 ~
R$\--matrix of $\B_n$ in terms of Jacobi theta functions.
Fortunately, we do not need the full details of this parametrization.
The essential thing for us is that the $\B_n$ statistical
weights depend on three
independent parameters:  the spectral parameter $z$, the anisotropy parameter
$\gamma$ and the modular parameter $\tau$.
As is customary, we treat $z$ as a variable and $\gamma, \tau$ as
parameters.
In fact it is convenient to treat $n$ as a parameter on equal
footing with $\gamma$ and $\tau$.
As mentioned above,
along with ${\cal B}_n$ we also find it useful to think in terms of the
corresponding
$(1+1)$\--dimensional field theoretical models ${\cal M}_n$.
The hamiltonian of ${\cal M}_n$ is obtained from the transfer matrix $T(z)$
of ${\cal B}_n$  through logarithmic differentiation at a special point.
We shall be interested in the antiferromagnetic regime with finite gap, so
that the ground state is constructed by filling the false (ferromagnetic)
vacuum with quasiparticles.
The partition function  of the ${\cal B}_n$ model in the thermodynamic
limit (the Perron\--Frobenius dominant eigenvalue of the transfer matrix)
was obtained by
Richey and Tracy [22].

A remarkable fact [23] in quantum integrable models is that the partition
function, as a function of the spectral variable $z$, coincides up to some
simple factors and/or redefinitions of parameters with a two\--particle
dressed $S$\--matrix, the spectral parameter acquiring the interpretation of
relative rapidity of the scattering particles.
We have to be more specific, there being $n-1$ types of dressed excitations
in ${\cal M}_n$.
We therefore briefly recall the picture of these excitations in the context
of the nested Bethe ansatz (BA).
The ground state, as was already mentioned is found by filling the false
vacuum with $n-1$ types of quasiparticles, each type at its own ``level''.
The momenta are distributed continuously in segments $ [-\pi , + \pi ]$ at
each level.
Excitations are viewed as ``holes'' in these distributions.
The type of physical excitation is determined by the level at which the hole
was created.
In terms of the system of interacting particles on the lattice associated to
the ${\cal M}_n$ model in the usual way, the first level corresponds to
charge excitations, while the others to ``isotopic'' excitations.
The levels are naturally ordered according to the sequence of the nested BA.
 From the $\B_n$ partition function of [19], [22], we extract
the (scalar)
$S$\--matrix $S_1^{(n)}$ for the scattering of  {\em two first} level
dressed
excitations
$$
S_1^{(n)} = \exp \left [
-i ( \frac{n-1}{n}) 2 \pi z - i F(z; \gamma; n , \frac{- i \pi \tau}{\gamma})
\right ]
\eqno(3a)
$$
where
$$
F(z; \gamma ; n , \frac{-i \pi \tau }{\gamma} ) = 2
 \sum_{k=1}^{\infty} {1 \over k}
\frac{{\rm sinh} ~ k \gamma (\frac{-i \pi \tau}{\gamma} - 1 ) }{{\rm sinh}
\frac{{\rm sinh}~ k \gamma (n-1)}{{\rm sinh}~k \gamma n } \cdot \sin 2 \pi k z
\eqno(3b)
$$
and the ``former'' spectral parameter $z$ which still appears in Eqs.~(3)
will presently be replaced by the relative rapidity $u$ (see Eq.~(5)).
$S_1^{(n)}$, as given in equations (3), differs slightly from the physical
$S$\--matrix for the scattering of two first level dressed excitations, by
some Blaschke-CDD pole factors and redefinitions of variables.
Yet $S_1^{(n)}$ shares all the essential features of the full $S$\--matrix
in question, while being much more convenient to work with.
So we will study this $S_1^{(n)}$.
The $S$\--matrix elements for the scattering of other types of excitations
are more complicated.
Therefore we restrict ourselves to the first level sector and its
$S$\--matrix $S_1^{(n)}$.

Notice the remarkable {\em symmetry} of $F(z; \gamma ; n , \frac{- i \pi
\tau}{\gamma })$ in its {\em last} two arguments
$$
F(z; \gamma ; n , \frac{-i \pi \tau}{\gamma}) = F (z; \gamma ; \frac{-i \pi
\tau}{\gamma}, n )~.
\eqno(4)
$$
Below
we shall get to make good use of the symmetry property (4).
Guided by Eq.~(4), we
introduce new variables
$$
\begin{array}{c}
l = \frac{i \gamma}{\pi \tau} \\
q = e^{i 2 \pi \tau} \\
u = - \frac{i z}{\tau} \\
\end{array}
\eqno(5)
$$
(our $q$ is the usual one, i.e. the square of the one in [22]).
Then we can write $e^{-iF}$ in the form
$$
\exp \left [
-iF(z; \gamma; n , \frac{- i \pi \tau}{\gamma}) \right ] =
\frac{\sigma (i u ; l ; n | q )}{\sigma (-i u ; l ; n | q )}
\eqno(6.a)
$$
with
$$
\sigma (i u ; l ; n | q ) = \prod_{k=0}^{\infty}
\frac{\Gamma_q (iu + nl (k+1))}{\Gamma_q (iu + nlk + l )}
\frac{\Gamma_q (iu + n l k + 1)}{\Gamma_q (iu + nlk + (n-1)l+1)} ~~~,
\eqno(6.b)
$$
and the $q$-gamma function $\Gamma_q(x)$ defined as [24]
$$
\Gamma_q(x) =
\frac{(q; q)_\infty}{(q^x, q)_\infty} (1-q)^{1-x} ~, ~~
(a, q)_\infty = \prod_{r=0}^{\infty} (1-aq^r ) ~~.
\eqno(7)
$$
Now let $n$ go to infinity.
Using the definitions (7) and keeping in mind that
as $x \air \infty$ for $q < 1$, we have
$(q^x, q)_\infty \air 1$, it is then readily seen that
$$
\sigma (iu ; l , \infty | q ) = [i u ]_q c (iu; l | q)
\eqno(8.a)
$$
where
$$[iu]_q =
\frac{1 - q^{iu}}{1-q}
\eqno(8.b)
$$
is the $q$-analogue of $i u$ and
$$
c (i u ; l | q ) =
\frac{\Gamma_q (i u )}{\Gamma_q (i u + l )}
\eqno(8c)
$$
Combining Eqs.~(8), (6a), (3a) we find
$$
S_1^{(n)} |_{n= \infty} = -
\frac{c (iu ; l | q )}{c ( - iu ; l | q ) }
\eqno(9)
$$
and we see the function $c ( i u ; l | q )$ emerging as the Jost function for
the scattering of first level dressed excitations of the $\M_\infty$ model.
Aside from the relative rapidity $u$, this function $c(iu; l | q)$ depends on
two parameters:  $l$ and $q$.
This eminently qualifies it as a candidate Harish\--Chandra $c$\--function
for the two parameter quantum symmetric space, which we are searching for.
Were this to be indeed the case, we ought to find pairs of values for the
parameters $l$ and $q$ for which the function $c (i u ; l | q)$ of Eq.~(8c)
reduces to the \p $c$\--functions $c_p$, Eq.~(1c) which we extracted from
the Mautner\--Cartier polynomials.
This is indeed the case.
If we introduce Macdonald's parameter
$$
t = q^l ~~,
\eqno(10)
$$
then for $q$ and $l$ {\em
both} going to zero in such a way
$$
t = q^l \air 1/p , ~~~~ q \air 0 , ~~~~ l \air 0 ~,
\eqno(11)
$$
we find that $c (i u; l | q )$ of Eq.~(8c) does reduce to $c_p$ of Eq.~(1c).
Note that for
$$
q \air 1~,~~~ ~~ l = 1/2~,~~~~~t = q^l \air 1 ~,
\eqno(12)
$$
$c (i u; l | q)$
becomes the Harish\--Chandra $c$\--function of the {\em
real} hyperbolic plane [16] $H_\infty = SL(2, \R )/ SO(2)$, the noncompact
partner of the 2\--sphere.
As we were, so far, expecting the 2\--sphere to turn up in this limit, one may
wonder how a Jost function (connected with a scattering process) could have
been obtained, how this switch to the noncompact partner could have
occurred.
We shall answer this question below, but first let us conclude the
argument.
If all this works, there must exist a family of orthogonal polynomials $P_n
(x; t |q )$ (here it is convenient to use $t$ in lieu of $l$,
(see Eq.~(10))) which provide
the corresponding zonal spherical functions, from which the function
$c (i u; l | q)$
can be extracted as Harish\--Chandra $c$\--function the same way the
$c_p$'s were extracted from the Mautner-Cartier polynomials.
But such a family of orthogonal polynomials,
as was already mentioned, is provided by the Macdonald
polynomials corresponding to root system $A_1$ [1],[2].
This family essentially coincides with the famous Roger\--Askey\--Ismail
(RAI) family [17], [18] of orthogonal polynomials.
Their asymptotics yield [15] the function $c (i u ; l | q )$ as
Harish\--Chandra $c$\--function.
As a check, these RAI polynomials reduce (up to some trivial
normalization factors) to the Mautner\--Cartier polynomials in the limit
(11), and to the Legendre polynomials of the
compact 2\--sphere in the limit (12).

At this point we can also account for the compact\--to\--non\--compact
switch which occurred when the $l = 1/2 ~~ q \air 1$ limit was taken at the
level of the $c$\--function.
The point is, that when
normalized like spherical functions, the RAI polynomials are {\em
symmetric} under the exchange
$$
n \leftrightarrow - (2 i \nu + l ) ~,
\eqno(13c)
$$
where $n$ is the order of the polynomial, $l$ is one of the deformation
parameters introduced earlier (Eq.~(10)), and $\nu$ is related to the
variable $\cos
\theta$ of the RAI polynomials and to the other deformation parameter $q$
by
$$
\nu = \theta / \log q
\eqno(13b)
$$
For $l = 1/2$, the case in question, the exchange (13) corresponds precisely
to the switch from the real 2\--sphere to the real hyperbolic plane.

In short then, from the  scattering of certain excitations in a problem endowed
with
an elliptic quantum algebra $Q_n , n \air \infty$, we have extracted a Jost
function which then turned out to be the Macdonald\--Harish\--Chandra
$c$\--function corresponding to the Macdonald polynomials for root system
$A_1$, the RAI polynomials.
We have thus {\em established a connection between the $Q_n$ algebras $(n
\air \infty )$ on one hand and these Macdonald polynomials on the other hand.}
This is our main result.

As was mentioned above, the data for the $Q_n$ algebra are    an elliptic
curve ${\cal E} = {\bf C /Z + Z} \tau_{\cal E}$,
characterized by the
modular parameter $\tau_{\cal E}$, or equivalently
$q_{\cal E} = \exp (i2 \pi
\tau_{\cal E})$,
and a point $\gamma$ on ${\cal E}$.
The data for the set of $A_1$\--Macdonald\--RAI polynomials are the two
parameters $t$ and $q$.
The connection between the elliptic and Macdonald parameters is then
$$
\begin{array}{c}
q = q_{\cal E} \\ q^l = t = e^{-2 \gamma}
\end{array}
\eqno(14)
$$
the second equation following directly from Eqs.~(5) and (10).
One may ask why we had to take the $n \air \infty$ limit.
On the face of it, all we should have had to deal with should have been the
elliptic algebra $ Q_2$
and the models which it underlies ${\cal B}_2$ and ${\cal M}_2$.
Going to $Q_n, {\cal B}_n , {\cal M}_n$ and then letting $ n \air \infty$ is
like searching for $SU(2)$ inside $SU(\infty )$.
For ordinary Lie groups this would be a detour, for elliptic quantum algebras
this may be needed on account of the complicated coproduct situation [14].
But once in $Q_\infty$, how is it we only found the Macdonald polynomials for
root system $A_1$ and not those for higher $A_n$ root systems?
The point is  we only looked at the scattering of
{\em two} first level excitations.

Whether or not the $n \air \infty$ limit is taken, it would be nice to have
a derivation of the Macdonald\--Harish\--Chandra $c$\--function of
Eq.~(8c) directly from symmetric spaces constructed from $Q_n$ quantum
groups,
along the standard lines of Harish\--Chandra theory (see e.g. [16]) and
without any reference to the physics of the $Z_n$\--Baxter models.
Conversely it would be of interest to find the geometric object for which
the  infinite product $\sigma (i u; l; n |q)$ of Eq.~(6b) is the
$c$\--function.
To steer all this into more familiar territory, notice that in the limit $q
\air 1$ the $q$\--gamma functions in the infinite product reduce to ordinary
gamma functions and the full infinite product (6a) becomes essentially
that which appears [25] in the soliton\--soliton scattering $S$\--matrix
in the
sine\--Gordon model, provided one
sets $n=2$ and relates our parameter $l$ to the
sine\--Gordon coupling constant $\beta$ via
$$
 l = \left ( \frac{8 \pi}{\beta^2} - 1 \right )~~.
\eqno(15)
$$
Thus the problem of understanding the ``geometric'' interpretation of the
infinite products has as an important special case soliton\--soliton scattering
in the sine\--Gordon model.
Conversely, we can regard the $S$\--matrix given by Eqs.~(3a), (5), (6) (7)
as a
``$q$\--deformation'' $\Gamma \air \Gamma_q$ of the sine\--Gordon soliton
scattering matrix of [26].
The Sklyanin algebra $(n=2)$ looks like the {\em further} deformation of
quantum $SL(2)$ by a new parameter.

With the just-established connection between ${\cal B}_n$
or ${\cal M}_n$ systems and
Macdonald polynomials it becomes interesting to see what happens
physics\--wise in the
regime in which the polynomials, ``go'' $p$\--adic.
For the $n \air \infty$ situation of Eq.~(9)  this corresponds to
$$
q=0 , ~~~~~~ t = e^{-2\gamma} = 1/p ~~.
\eqno(16a)
$$
so that
$$
\gamma = \log \sqrt{p}
\eqno(16b)
$$
Eqs.~(9), (5), (8c) and (7) then yield
$$
S_1^{(n)} \left |_{n= \infty, q=0, t=1/p} =
\frac{pe^{i2 \pi z} - 1}{e^{i 2 \pi z} - p} ~,
\right .
\eqno(17)
$$
which coincides with the {\em bare} $S$\--matrix in the XXZ model with the
same value of $\gamma$.
Could this result also be obtained from a model on a Bethe lattice with
$p+1$ edges incident at each vertex?

A direct study of ${\cal M}_\infty$ models using the powerful quantum
inverse scattering method or the Bethe ansatz is highly desirable.

The other interesting case is
$$
t = q^{1/2} ~~,
\eqno(18a)
$$
so that
$$
l = 1/2 ~~.
\eqno(18b)
$$
This corresponds to the familiar {\em
one}\--parameter quantum group $SU(2)_q$.
In the limit $q \air 1$ it then yields the ordinary Lie group $SU(2)$.
For the $SU(2)_q$ case we have the direct treatment by one of us [9].
An immediate comparison with the results of [9] is not possible, since
there $n = 2$, whereas for us
$$
n \air \infty ~, ~~~~~~~~ l^{-1} = -
\frac{i \pi \tau}{\gamma} = 2
\eqno(19)
$$
It is clear from Eq.~(3a) that for large $n$, the $S$\--matrix depends only
on the function $F(z; \gamma ; n ; \frac{i \pi \tau}{\gamma} )$.
But, as we saw in Eq.~(4) this function is symmetric under the interchange
of its last two arguments.
Therefore instead of the case (19) we can deal with the equivalent case of
$$
n = 2 , ~~~~~~~
\frac{-i \pi \tau}{\gamma} \air  \infty
\eqno(20)
$$
which is then in line with [9], provided one replaces (5) by
$$
l = \frac{1}{n} ~, ~~~ q = e^{-2 \gamma n} ~~~ u = \frac{\pi
z}{\gamma n} ~~,
\eqno(21)
$$
so that, yet again
$$
t = q^l = e^{-2 \gamma} ~~.
\eqno(22)
$$
As in [12], we find the XXZ model in this case.
This is also in line with the connection between the XXZ\--model and
$SU(2)_q$  displayed in [26].

We can also view the \p and $SU(2)_q$ cases directly on the ${\cal B}_2$ or
8\--vertex model or on the related XYZ model
by using the more involved formula for $S_1^{(n)}$ or for the partition
function for $\B_2$.
In terms of Baxter's parameters [17] the \p case corresponds to the
6\--vertex model in the
principal antiferroelectric regime with
$$
\Gamma = 1,~~~~ ~ \Delta = -
\frac{p^{1/2} + p^{- 1/2}}{2},
\eqno(23)
$$
In terms of the XXZ chain this corresponds to the antiferromagnetic XXZ
chain
$(\Gamma = 1)$ with asymmetry $\Delta$ given by (21) (remember $J_x  : J_y
: J_z = 1:
\Gamma : \Delta $).

Similarly the case $t = q^{1/2}$ of the ordinary one\--parameter quantum
group $SU(2)_q$  corresponds to
$$
\Gamma = 0 ~~~~~~~
\Delta = - \frac{1+k}{2 \sqrt{k}}
\eqno(24)
$$
where $k$ is the modulus of the elliptic Jacobi functions of nome $q$.
This is the XZ model.
If we now let $t = q^{1/2} \air 1$, corresponding to the ordinary $SU(2)$
Lie group,
then the comodulus $k' \air 0$, so that the modulus $k \air 1$ and $\Delta
\air - 1$.

A large class of elliptic quantum algebras has already been brought into
play in the context of the $Z_n$-Baxter models and the simplest Macdonald
polynomials.
The question then naturally arises as to a full classification of
elliptic quantum
symmetric spaces, in correspondence with admissible pairs of root
systems.
Moreover the
parameters $q,t$ of Macdonald translate on the ``Sklyanin side'' into an
elliptic curve and a point on it.  Could one make the connection with
elliptic curves explicit directly on the ``Macdonald side''?

For generic $t$ and $q$, the orthogonal RAI polynomials obey of course,
a three term recursion relation.
In the \p regime $(q=0, ~ t = 1/p)$ this recursion relation becomes precisely
the condition that the zsf's be eigenfunctions of the laplacian.
On the Cayley\--Bethe\--Bruhat\--Tits tree, corresponding to this case, the
laplacian has a
simple interpretation as a difference operator obeying the mean value
theorem.
It is then natural to expect that
for generic $t$ and $q$,
the recursion relation (3.6)
also corresponds to the requirement that the RAI
polynomials be eigenfunctions of the laplacian on some ``non\--arboreal''
{\em discrete} space, which reduces to a tree in the \p regime.
It would be nice to find a simple geometric description of this generic
discrete space which in the \p case becomes a tree whereas for $t = q^{1/2}
\air 1$ becomes a sphere of (real) dimension 2.
In short then, it would be interesting to have a direct geometric picture of
the quantum symmetric  space, not only of its zonal spherical functions.

In this paper we have dealt only with two\--body scattering.
For $N$\--body scattering one can construct a factorized $S$\--matrix, the
corresponding factorization formulae being quantum analogues of the familiar
Gindikin\--Karpelevi\u{c} formulae [16].
Although Gindikin\--Karpelevi\u{c} factorization formulae for higher rank
quantum symmetric spaces have not (to our knowledge) been proved to date,
our results indicate -- not surprisingly -- that such formulae ought to exist.

In addition to the connection, discussed here,
between real and \p quantum\--symmetric spaces, there exists a further known
connection via adelic constructs [27]-[29].
This raises the possibility of combining these two types of real\--to\--\p
connection.
This has led us to the study of $q$\--deformed (or more
accurately $l$\--deformed) Euler products.
At the root of such ideas lies the observation that in a well defined sense,
the $q$\--exponential
$e_q (-x^{1/l})$ plays the role of a ``quantum gaussian'' which
interpolates between the familiar real gaussian $(\exp
(-x^2))$ and the \p ``gaussian'' i.e. the characteristic function of the set
$\Z_p$ of \p integers.
We obviously use the term gaussian with the meaning of Fourier
self\--similar function.
Here we shall not go further into these topics, but refer the reader to the
expanded version of this paper [15].

We still wish to draw attention to the fact that our work bears on string
theory.
We have constructed {\em two\--parameter} deformations of string theory
which ``interpolate'' between \p and ordinary Veneziano strings, and when a
relation of the type $t = q^{1/2}$ is imposed between the two parameters
they yield ``$q$\--strings''.
This will be detailed elsewhere.

\bigskip

We wish to thank S. Bloch, L. Chekhov, L. Mezincescu, A. D. Mironov and M.
Olshanetsky.

\bigskip
\centerline{REFERENCES}

\begin{enumerate}

\item I.~G.~Macdonald,  in {\em Orthogonal Polynomials:
Theory and Practice}, P.~Nevai ed., Kluwer Academic Publ., Dordrecht, 1990.

\item I.~G.~Macdonald,  Queen Mary College preprint 1989.

\item P.~G.~O.~Freund,  in {\em Superstrings and Particle Theory},
L.~Clavelli and B.~Harms eds., World Scientific, Singapore, 1990.

\item P.~G.~O.~Freund,
in {\em Quarks, Symmetries and Strings, A Symposium in Honor of Bunji
Sakita's 60th Birthday}, M.~Kaku, A.~Jevicki and K.~Kikkawa eds., World
Scientific, Singapore, 1991.

\item A.~V.~Zabrodin,  Moscow preprint 1991,
Mod. Phys. Lett. A, in press.

\item P.~G.~O.~Freund,
Phys. Lett. {\bf 257B} (1991) 119.

\item F.~Mautner,  Am. J. Math. {\bf 80} (1958) 441.

\item P.~Cartier,  {\em Proc. Symp. Pure Math.} Vol. {\bf 26},
A.~M.~S.~Providence, 1973.

\item A.~V.~Zabrodin,  Comm. Math. Phys. {\bf 123} (1989) 463.

\item E.~K.~Sklyanin,  Funk. Anal. Appl. {\bf 16} (1982) 263.

\item E.~K.~ Sklyanin,  Funk. Anal. Appl. {\bf 17} (1983) 273.

\item I.~V.~Cherednik,  Yad. Fiz. {\bf 36} (1982) 549.

\item I.~V.~Cherednik,
Funk. Anal. Appl. {\bf 19} (1985) 77.

\item A.~V.~Odeskii,  and B.~L.~Feigin,  Funk. Anal. Appl. {\bf 23}
(1989) 207.

\item P.~G.~O.~Freund,  and A.~V.~Zabrodin, Chicago preprint EFI 91-43.

\item S.~Helgason, {\em Topics in Harmonic Analysis on Homogenous Spaces},
Birk\-h\"{a}user, Basel, 1981

\item L.~J.~Rogers,  Proc. London Math. Soc. {\bf 26} (1895) 318.

\item R.~Askey,  and R.~E.~H.~Ismail,  in {\em Studies in Pure Math.}
P.~Erd\"{o}s, ed.
Birkh\"{a}user, Basel, 1983.

\item R.~J.~Baxter,  {\em Exactly Solved models in Statistical
Mechanics},
Acad. Press, N.Y., 1982

\item A.~A.~Belavin,  Nucl. Phys. {\bf B180} 189 (1981).

\item P.~P.~Kulish,  and N.~Yu.~Reshetikhin,  Soviet Phys. JETP {\bf
53} (1981) 108 (1981).

\item M.~P.~Richey,  and C.~A.~Tracy,  J. Stat. Phys. {\bf 42} (1986) 311.

\item A.~B.~Zamolodchikov,
Comm. Math. Phys. {\bf 69} (1979) 165.

\item G.~Gasper,  and M.~Rahman, {\em Basic Hypergeometric Series},
Cambridge
Univ. Press, Cambridge, 1990

\item A.~B.~Zamolodchikov,  and
Al.~B.~Zamolodchikov,  Ann. Phys. (N.Y.) {\bf 120} (1979) 253.

\item V.~Pasquier and H. Saleur, Nucl. Phys. {\bf B330} (1990) 523.

\item J.~Tate,  Thesis, Princeton 1950, reprinted in {\em Algebraic Number
Theory}, J.~W.~S.~Cassels and A.~ Fr\"{o}hlich, eds., Academic Press, N.Y.,
1967.

\item I.~M.~Gel'fand,  M.~I.~Graev,  and I.~I.~Pyatetskii\--Shapiro,
{\em
Representation Theory and Automorphic Functions}, Saunders, London, 1966.

\item A.~Weil, {\em Adeles and Algebraic Groups} Birkh\"{a}user, Basel 1982.

\end{enumerate}

\end{document}